# Hybrid Approaches to Image Coding: A Review

Rehna. V. J
Department of Electronics & Communication Engineering
Noorul Islam University
Kanyakumari District, India
rehna_vj@yahoo.co.in

Jeya Kumar. M. K
Department of Computer Application
Noorul Islam University
Kanyakumari District, India
jeyakumarmk@yahoo.com

*Abstract*— **Now a days, the digital world is most focused on storage space and speed. With the growing demand for better bandwidth utilization, efficient image data compression techniques have emerged as an important factor for image data transmission and storage. To date, different approaches to image compression have been developed like the classical predictive coding, popular transform coding and vector quantization. Several second generation coding schemes or the segmentation based schemes are also gaining popularity. Practically efficient compression systems based on hybrid coding which combines the advantages of different traditional methods of image coding have also being developed over the years. In this paper, different hybrid approaches to image compression are discussed. Hybrid coding of images, in this context, deals with combining two or more traditional approaches to enhance the individual methods and achieve better quality reconstructed images with higher compression ratio. Literature on hybrid techniques of image coding over the past years is also reviewed. An attempt is made to highlight the neuro-wavelet approach for enhancing coding efficiency.**

## I. INTRODUCTION

Image Compression is an important area in the field of digital image processing. It deals with techniques for reducing the storage space required for saving an image or the bandwidth required for transmitting it. The main goal of image compression is to reduce irrelevance and redundancy of the image data thereby optimizing the storage space and increasing the transmission rate over WebPages. Image compression is performed in such a way that it enables image reconstruction. The amount of compression achieved depends on the contents of image data. A typical photographic image can be compressed to about 80% of its original size without experiencing noticeable degradation in the quality. The strong correlation between image data items enables reduction in data contents without significant quality degradation.

Image compression schemes [1] are generally classified as lossless compression schemes and lossy compression schemes. Lossless compression is an error free compression where the original data can be recovered after decompression. This scheme provides low compression ratio but has several applications, like in the compression of medical images where the loss of information are not acceptable. In lossy compression, some extend of the original data is lost during compression and so, after the decompression process, an approximation of original data is obtained.

Lossy schemes achieves higher compression ratio than lossless schemes and are used in applications, like compression of natural images where perfect reconstruction is not essential and we can afford the partial loss in the image data as long as it is within tolerance. During the design of a lossy compression scheme, two important issues are considered, the compression ratio and the tolerance in the visual quality degradation. Compression ratio [2] is the ratio of the size of the original image to that of the compressed image. It gives an indication of how much compression is achieved for a particular image. The compression ratio, starting at 1 with the first digital picture in the early 1960s, has reached a saturation level around 300:1 a couple of years ago. Image quality still remains as an important problem to be investigated. As the compression ratio increases, the quality of the resulting image degrades. So, a trade off between compression ratio and the tolerance in the visual quality degradation need to be considered during compression.

This paper focuses on hybrid image coding systems which combine the advantages of different classical methods to enhance the individual techniques and achieve better quality reconstructed image with higher compression ratio. Stress is given on the combined approach of image compression using neural networks and wavelet transforms, called the neuro-wavelet approach.

Rest of the paper is organized as follows: Section II deals with the error matrics, section III briefs about traditional techniques of image coding, section IV gives an overview of the early literature on image compression, section V is devoted to hybrid techniques to image compression followed by VQ-based hybrid techniques in section V.1, wavelet-based hybrid techniques in section V.2, ANN-based hybrid techniques in V.3 and the Neuro-Wavelet model in section V.4. Discussion & conclusion of the review is presented in section VI.

## II. ERROR MATRICS

The different compression algorithms can be compared based on certain performance measures. Compression Ratio (CR) is the ratio of the number of bits required to represent the data before compression to the number of bits required after compression. Bit rate is the average number of bits per sample or pixel (bpp), in the case of image. The image quality can be evaluated objectively and subjectively. A standard objective measure of image quality is reconstruction error given by equation 1.

$$\text{Error, E} = \text{Original image} - \text{Reconstructed image} \qquad (1)$$

Two of the error metrics used to compare the various image compression techniques are the mean square error (MSE) and the Peak Signal to Noise Ratio (PSNR). MSE refers to the average value of the square of the error between the original signal and the reconstruction as given by equation 2. The important parameter that indicates the quality of the reconstruction is the peak signal-to-noise ratio (PSNR). PSNR is defined as the ratio of square of the peak value of the signal to the mean square error, expressed in decibels.

$$MSE = \text{E} / (\text{SIZE OF IMAGE}) \qquad (2)$$

The MSE is the cumulative squared error between the compressed and the original image, whereas PSNR is a measure of the peak error. The mathematical formulae for the computation of MSE & PSNR is :

$$MSE = 1/MN \left[\sum_{i=1}^{M} \sum_{j=1}^{N} (I_{ij} - I'_{ij})^2\right] \qquad (3)$$

$$PSNR = 20 * \log 10\ (255 / \text{sqrt}(MSE)) \qquad (4)$$

where I(x,y) is the original image, I'(x,y) is the approximated version (which is actually the decompressed image) and M, N are the dimensions of the images, 255 is the peak signal value. A lower value for MSE means lesser error, and as seen from the inverse relation between the MSE and PSNR. Higher values of PSNR produce better image compression because it means that the ratio of Signal to Noise is higher. Here, the 'signal' is the original image, and the 'noise' is the error in reconstruction. So, a compression scheme having a lower MSE (and a high PSNR), can be recognized as a better one. Subjective quality is measured by psychophysical tests and questionnaires.

## III. TRADITIONAL TECHNIQUES

To date, many compression algorithms have been developed for image coding such as the classical predictive coding[3], the popular transform coding [4], the commercially successful wavelet coding [5] and vector quantization [6]. Predictive coding refers to the de-correlation of similar neighbouring pixels within an image to remove redundancy. Transform coding, an efficient coding scheme is based on utilization of inter-pixel correlation which is a core technique recommended by JPEG. Wavelet coding is a popular form of data compression well suited for image and audio compression. The wavelet transform have become the most prevalent techniques among the image coding techniques as they are localized in both spatial and frequency domains. With wavelets, a compression rate of up to 1:300 can be achieved. Wavelet compression allows the integration of various compression techniques into one algorithm. Vector quantization, a technique often used in lossy data compression requires the development of an appropriate codebook to compress data.

Another recent lossy image compression method is the fractal compression [7, 32] which is best suited for textures and natural images, relying on the fact that parts of an image often resemble other parts of the same image. Fractal algorithms convert these parts into mathematical data called "fractal codes" which are used to recreate the encoded image. Recently, several segmentation based, or the second generation image coding techniques [8] are also gaining popularity. Many image compression algorithms such as the Bandelets [9], the Prune tree [10], the Prune-Join tree [10], the GW image coding method [11] based on the sparse geometric representation etc have also been introduced. Image compression is also achieved with considerable efficiency using neural networks [12, 13] due to their parallel architectures and flexibility. Neural networks have the ability to preprocess input patterns to produce simpler patterns with fewer components. This compressed information (stored in a hidden layer) preserves the full information obtained from the external environment.

## IV. EARLY LITERATURE

The concept of compressing 2 dimensional signals, especially images was introduced in the year 1961, by Wholey. J [14]. The first data compression approach was the predictive coding technique in which the statistical information of the input data is considered to reduce redundancy. Although the resulting compression was not great, there were reasons for believing that this procedure would be more successful with realistic pictorial data. A method of data compression by run length encoding was published in 1969, by Bradley, S.D[15]. In his paper, the optimal performance of the code was for a compression factor of 13.20. An adaptive variable length coding system was presented by Rice et al. in 1971 [16]. Using sample to sample prediction, the coding system produces output rates within 0.25 bits/picture element (pixel) of the one dimensional difference entropy, for entropy values ranging from 0 to 8 bits/pixel.

The transform coding approaches to image compression was introduced in the year 1971 with application of discrete Fourier transform for achieving image compression [17]. Pratt and Andrews [18] studied bandwidth compression using the Fourier transform of complete pictures. The most commonly used transform coding uses the Fourier related transforms such as the KL transform [19], HADAMARD transform [20] etc. Singular value decomposition [21] is the representation of data using smaller number of variables and had been widely used for face detection and object recognition. SVD has been applied for image compression in 1976 and was found successful. In the past decades, Discrete Cosine Transform or DCT has been the most popular for image coding because it provides optimal performance and can be implemented at a reasonable cost.

Recent advances in signal processing tools such as wavelets opened up a new horizon in sub and image coding. Wavelets were applied to image coding in the year 1989. The wavelet transform have proven to be very effective and has gained popularity over the DCT. Discrete Wavelet Transforms has the ability to solve the blocking effect introduced by the DCT. Different wavelets and its variants were used in later years to achieve better compression. These include the harmonic wavelets, the very efficient embedded zero tree wavelet (EZW) [22], the much popular SPIHT (Set Portioning in Hierarchical Trees) [23], EBCOT (Embedded Block Coding with Optimized Truncation) [24] etc to name a few. EZW coding for image compression presented by Shapiro in his paper "Smart compression using the EZW algorithm" in 1993 uses wavelet coefficients for coding. EZW coding exploits the multi resolution properties of the wavelet transforms to give a computationally simple algorithm with better performance compared to other existing wavelet transforms. The techniques of EZW, SPIHT, EBCOT etc are used as reference for comparison with the new techniques of image compression.

The second generation coding methods [8] or the segmentation based methods of image coding where introduced in the year 1985 and many variations have been introduced since then. Two groups can be formed in this class: methods using local operators and combining their output in a suitable way and methods using contour-texture descriptions. For low bit-rate compression applications, segmentation-based coding methods provide, in general, high compression ratios when compared with traditional (e.g. transform and subband) coding approaches. Among the many segmentation techniques that have been developed, the binary space partition scheme (BSP) [25] is a simple and effective method of image compression. The concept of BSP for hidden surface removal was published in the year 1996 by Hyder Radha et. al. Recently the BSP scheme is enhanced with the wavelet approach called as geometric wavelets [11] by Dekel et. al.

Over the last decade, numerous attempts have been made to apply artificial neural networks (ANNs) for image compression. Neural network approaches used for data compression seem to be very efficient due to their structures which offer parallel processing of data. Kohenen's self organizing maps were the first neural networks used to achieve vector quantization for image compression. This is discussed by Luttrell, S.P [26] in his paper. Direct applications of neural networks to image compression came with the Back Propagation Neural Network (BPN) [29] in 1989 proposed by Sonehara et al. Apart from Kohonen Selforganizing Maps (SOM) and BPNs, other neural networks that are used for image coding include Hierarchical SOMs (HSOMs)[27], Modular Neural networks (MNNs)[30], and Cellular Neural Networks[28]. Researches on neural networks for image compression are still making steady advances.

## V. HYBRID TECHNIQUES

Hybrid approaches to image compression deals with combining two or more traditional approaches to enhance individual methods and achieve better quality reconstructed images with higher compression ratio. The concept of hybrid coding was introduced in the early 1980's itself by Clarke R.J [31] on transform coding combined with predictive coding. Since then, various hybrid techniques have evolved namely the vector quantization combined with DCT, Block Truncation Coding (BTC) with Hopfield neural networks, predictive coding with neural networks wavelet coding with neural networks, segmentation based techniques with predictive coding techniques, fractal coding with neural networks, subband coding with arithmetic coding, segmentation coding with wavelet coding, DCT with DWT, SPIHT with fractal coding, cellular neural networks with wavelets etc. Some of the different successful hybrid coding techniques are discussed in the next few sections.

### *V.1 VQ-Based Hybrid Techniques*

Vector quantization (VQ) is a successful, effective, efficient, secure, and widely used compression technique over two decades. The strength of it lies in higher compression ratio and simplicity in implementation, especially of the decoder. The major drawback of VQ is that, decompressed image contains blockiness because of the loss of edges due to which the quality of image degrades. VQ has been combined with traditional techniques of image coding to achieve better performance compared to the conventional VQ scheme. Some works related to VQ based hybrid approaches to image coding over the past two decades are discussed here.

Vector Quantization of images based on a neural network clustering algorithm, namely the Kohenons Self Organising Maps proposed by Feng et. Al [33] were one of the early works in hybrid VQ techniques. P. Daubechies et. al introduced the coding of images using vector quantization in the wavelet transform domain [34]. Here, a wavelet transform is first used in order to obtain a set of orthonormal subclasses of images; the original image is decomposed at different scales using pyramidal algorithm architecture. The decomposition is along the vertical and horizontal directions and maintains the number of pixels required to describe the image at a constant. Then according to Shannon's rate-distortion theory, the wavelet coefficients are vector quantized using a multiresolution codebook. To encode the wavelet coefficients, a noise-shaping bit-allocation procedure which assumes that details at high resolution are less visible to the human eye is proposed in the work.

A hybrid BTC-VQ-DCT (Block Truncation Coding-Vector Quantization - Discrete Cosine Transform) [35] image coding algorithm was proposed by Wu et. al. The algorithm combines the simple computation and edge preservation properties of BTC and the high fidelity and high-compression ratio of adaptive DCT with the high-compression ratio and good subjective performance of VQ. This algorithm and can be implemented with significantly lower coding delays than either VQ or DCT alone. The bit-map generated by BTC is decomposed into a set of vectors which are vector quantized. Since the space of the BTC bit-map is much smaller than that of the original 8-bit image, a lookup-table-based VQ encoder has been designed to `fast encode' the bit-map. Adaptive DCT coding using residual error feedback is implemented to encode the high-mean and low-mean subimages. The overall computational complexity of BTC-VQ-DCT coding is much less than either DCT or VQ, while the fidelity performance is competitive. The algorithm has strong edge-preserving ability because of the implementation of BTC as a precompress decimation. The total compression ratio achieved is about 10:1.

A hybrid coding system that uses a combination of set partition in hierarchical trees (SPIHT) and vector quantisation (VQ) for image compression was presented by Hsin. H. C et. Al [36]. Here, the wavelet coefficients of the input image are rearranged to form the wavelet trees that are composed of the corresponding wavelet coefficients from all the subbands of the same orientation. A simple tree classifier has been proposed to group wavelet trees into two classes based on the amplitude distribution. Each class of wavelet trees is encoded using an appropriate procedure, specifically either SPIHT or VQ. Experimental results show that advantages obtained by combining the superior coding performance of VQ and efficient cross-subband prediction of SPIHT are appreciable for the compression task, especially for natural images with large portions of textures.

Arup Kumar Pal et. al have recently proposed a hybrid DCT-VQ based approach for efficient compression of color images [37]. Initially DCT is applied to generate a common codebook with larger codeword sizes. This reduces computation cost and minimizes blocking artifact effect. Then VQ is applied for final compression of the images that increases PSNR. Table 1 shows the simulation results for two test images using the conventional VQ method and the proposed hybid method. Better PSNR values are obtained for the hybrid technique compared to the conventional VQ process. So the proposed hybrid scheme improves the visual quality of the reconstructed image compare to the conventional VQ process. The simulation result also shows that the proposed scheme reduces the computation cost (including codebook construction, VQ encoding process and VQ decoding process time) compared to the conventional VQ process.

TABLE I. SIMULATION RESULTS SHOWING THE COMPARISON BETWEEN CONVENTIONAL VQ & HYBRID VQ

| **Method** | **Code-book Size** | **Block Size** | **Test Images (512 x 512)** | | | |
|---|---|---|---|---|---|---|
| | | | **Sailboat** | | **Pepper** | |
| | | | ***PSNR in dB*** | ***Overall Computation Time in seconds*** | ***PSNR in dB*** | ***Overall Computation Time in seconds*** |
| ***Conven-tional VQ*** | 1024 | 8 x 8 | 25.6392 | 485.91 | 27.6969 | 484.01 |
| | | 16 x 16 | 23.5795 | 520.02 | 25.5791 | 518.69 |
| ***Hybrid DCT-VQ*** | 1024 | 8 x 8 | 26.7528 | 164.34 | 28.6675 | 159.75 |
| | | 16 x 16 | 25.3996 | 43.35 | 26.3590 | 42.34 |

These works are some examples that prove the improvement in performance of VQ based hybrid techniques over the individual methods.

### *V.2 Wavelet Based Hybrid Techniques*

The wavelet transform, which provides a multiresolution representation of images, has been widely used in image compression. Wavelet transforms have been combined with classical methods of image coding to obtain high quality compressed images with higher compression ratios. Some of the wavelet based hybrid techniques are discussed in this section. Durrani et. Al [38] combined the run length encoding with the wavelet transforms to achieve better compression. The main attraction of this coding scheme is its simplicity in which no training and storage of codebooks are required. Also its high visual quality at high compression ratio outperforms the standard JPEG codec for low bitrate applications. Jin Li Kuo et. al [39] proposed a hybrid wavelet-fractal coder (WFC) for image compression. The WFC uses the fractal contractive mapping to predict the wavelet coefficients of the higher resolution from those of the lower resolution and then encode the prediction residue with a bitplane wavelet coder. The fractal prediction is adaptively applied only to regions where the rate saving offered by fractal prediction justifies its overhead. A rate-distortion criterion is derived to evaluate the fractal rate saving and used to select the optimal fractal parameter set for WFC. The superior performance of the WFC is demonstrated with extensive experimental results.

According to F. Madeiro et. al., [40] wavelet based VQ is the best way of quantizing and compressing images. This methodology takes multiple stage discrete wavelet transform of code words and uses them in both search and design processes for the image compression. Accordingly, the codebook consists of a table, which includes only the wavelet coefficients. The key idea in the mechanism of this algorithm is finding representative code vectors for each stage, They are found by first combining n code words in k groups, where kn gives the codebook size. This technique has a major drawback in the amount of computations during the search for optimum code vector in encoding. This complexity can be reduced by using an efficient codebook design and wavelet based tree structure.

Iano, Y et. Al [41] presents a new fast and efficient image coder that applies the speed of the wavelet transform to the image quality of the fractal compression. Fast fractal encoding using Fisher's domain classification is applied to the lowpass subband of wavelet transformed image and a modified set partitioning in hierarchical trees (SPIHT) coding, on the remaining coefficients. Furthermore, image details and wavelet progressive transmission characteristics are maintained, no blocking effects from fractal techniques are introduced, and the encoding fidelity problem common in fractal-wavelet hybrid coders is solved. The proposed scheme promotes an average of 94% reduction in encoding-decoding time comparing to the pure accelerated Fractal coding results. The simulations results show that, the new scheme improves the subjective quality of pictures for high-medium-low bitrates.

Alani et. al [11] proposes a well suited algorithm for low bit rate image coding called the Geometric Wavelets. Geometric wavelet is a recent development in the field of multivariate piecewise polynomial approximation. Here the binary space partition scheme which is a segmentation based technique of image coding is combined with the wavelet technique [51]. The discrete wavelet transforms have the ability to solve the blocking effect introduced by the DCT. They also reduce the correlation between neighboring pixels and gives multi scale sparse representation of the image. Wavelet based techniques provide excellent results in terms of rate distortion compression, but they do not take advantage of underlined geometry of the edge singularities in an image. The second generation coding techniques exploits the geometry of the edge singularities of the image. Among them the binary space partition scheme is a simple and efficient method of image coding, which is combined with geometric wavelet tree approximation so as to efficiently capture edge singularities and provide a sparse representation of the image. The geometric wavelet method successfully competes with the state-of-the-art wavelet methods such as EZW, SPIHT and EBCOT algorithms. A gain of 0.4 dB over SPIHT and EBCOT algorithms is reported. This method also outperforms other recent methods that are based on sparse geometric representation. For eg: this algorithm reports a gain of 0.27 dB over the bandelets algorithms at 0.1 bits per pixel.

A hybrid compression method for integral images using Discrete Wavelet Transform and Discrete Cosine Transform is proposed by Elharar et. Al [42]. A compression method is developed for the particular characteristics of the digitally recorded integral image. The compression algorithm is based on a hybrid technique implementing a four-dimensional transform combining the discrete wavelet transform and the discrete cosine transform. The proposed algorithm outperforms the baseline JPEG compression scheme.

### *V.3 ANN Based Hybrid Techniques*

The existing conventional image compression technology can be developed into various learning algorithms to build up neural networks for image compression. This will be a significant development and a wide research area in the sense that various existing image compression algorithms can actually be implemented by one neural network architecture empowered with different learning algorithms.

Hence, the powerful parallel computing and learning capability of neural networks can be fully exploited to build up a universal test bed where various compression algorithms can be evaluated and assessed. Three conventional techniques are covered in this section, which include predictive coding, fractal coding, and wavelet transforms.

### *A. Predictive Coding Neural Networks*

Predictive coding has been proved to be a powerful technique in de-correlating input data for speech and image compression where a high degree of correlation is embedded among neighboring data samples. The autoregressive (AR) model, a classification of predictive coding, has been successfully applied to image compression. Predictive coding in terms of applications in image compression can be further classified into linear and non-linear AR models. Conventional technology provides a mature environment and well developed theory for predictive coding which is represented by LPC (linear predictive coding), PCM (pulse code modulation), DPCM (delta PCM) or their modified variations. Non-linear predictive coding, however, is very limited due to the difficulties involved in optimizing the coefficients extraction to obtain the best possible predictive values. Under this circumstance, a neural network provides a very promising approach in optimizing non-linear predictive coding [43, 44]. Based on a linear AR model, a multilayer perceptron neural network can be constructed to achieve the design of its corresponding non-linear predictor as shown in Fig. 1. For the pixel $X_n$ which is to be predicted, its $N$ neighbouring pixels obtained from its predictive pattern are arranged into a one dimensional input vector $X=\{X_{n-1}, X_{n-2},\ldots,X_{n-N}\}$ for the neural network.

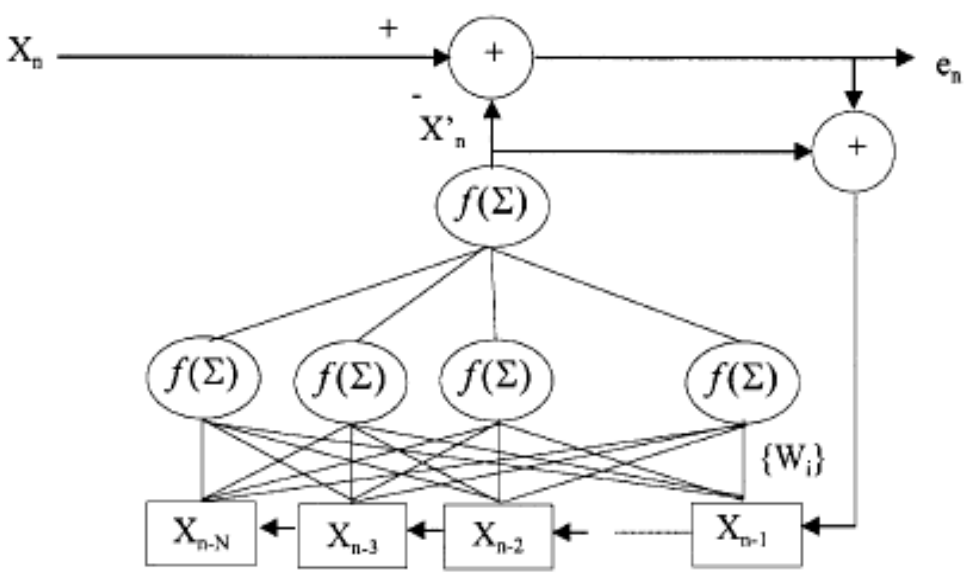

Figure 1 Predictive Neural Network

A hidden layer is designed to carry out back propagation learning for training the neural network. Predictive performance with neural networks is claimed to outperform the conventional optimum linear predictors by about 4.17 and 3.74 dB for two test images [44]. Further research, especially for non-linear networks, is encouraged by the reported results to optimize their learning rules for prediction of those images whose contents are subject to abrupt statistical changes.

### *B. Fractal Neural Networks*

Fractal configured neural networks [45,46], based on iterated function system (IFS) codes [47], represent another example along the direction of combining existing image compression technology with neural networks. Its conventional counterpart involves representing images by fractals and each fractal is then represented by so called IFS, which consists of a group of affined transformations. To generate images from IFS, random iteration algorithm is used which is the most typical technique associated with fractal based image decompression [47]. Hence, fractal based image compression features lower speed in compression and higher speed in decompression. By establishing one neuron per pixel, two traditional algorithms of generating images using IFSs are formulated into neural networks in which all the neurons are organized as a topology with two dimensions [46].

Fig. 2 illustrates the network structure in which $w_{ij,i'j'}$ is the coupling weight between $(ij)$th neuron to $(i'j')$th one, and $s_{ij}$ is the state output of the neuron at position $(i,j)$. The training algorithm is directly obtained from the random iteration algorithm in which the coupling weights are used to interpret the self similarity between pixels. In common with most neural networks, the majority of the work operated in the neural network is to compute and optimize the coupling weights, $w_{ij,i'j'}$. Once these have been calculated, the required image can typically be generated in a small number of iterations. Hence, the neural network implementation of the IFS based image coding system could lead to massively parallel implementation on a dedicated hardware for generating IFS fractals. Although the essential algorithm stays the same as its conventional algorithm, solutions could be provided by neural networks for the computing intensive problems, which are currently under intensive investigation in the conventional fractal based image compression research area.

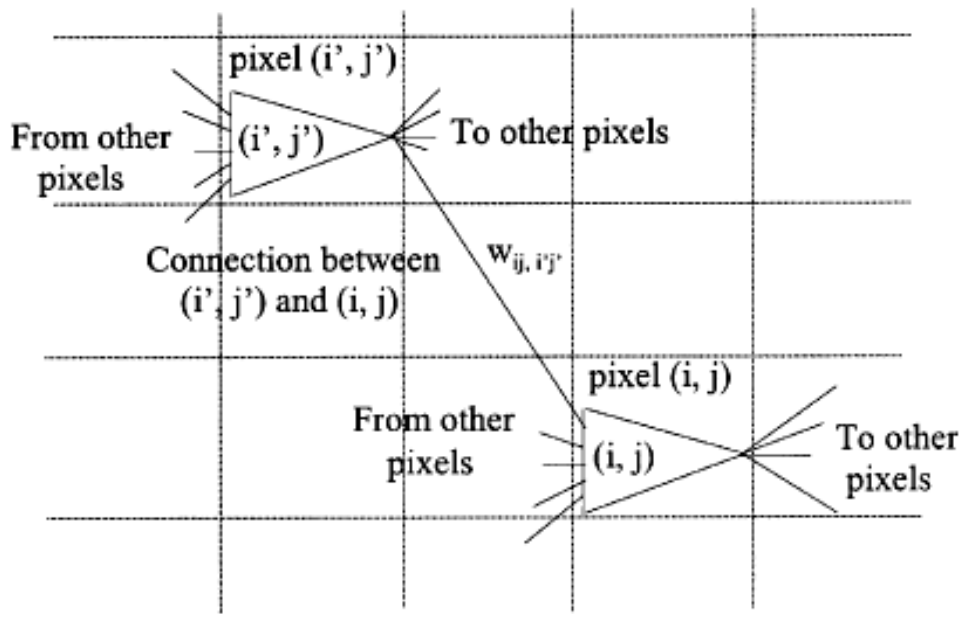

Figure 2 Fractal Neural network

## C. *Wavelet Neural Networks*

Based on wavelet transforms, a number of neural networks are designed for image processing and representation [48, 49]. Wavelet networks are a combination of radial basis function (RBF) networks and wavelet decomposition, where radial basis functions were replaced by wavelets. When a signal $s$(t) is approximated by daughters of a mother wavelet $h(t)$, for instance, a neural network structure can be established as shown in Fig. 3 [48, 49]. Here, step1 computes a search direction [$\boldsymbol{s}$] at iteration $i$. Step2 computes the new weight vector using a variable step-size α. By simply choosing the stepsize α, as the learning rate, the above two steps can be constructed as a learning algorithm for the wavelet neural network in Fig.3. Experiments reported [19] on a number of image samples support the wavelet neural network by finding out that Daubechie's wavelet produces a satisfactory compression with the smallest errors. Haar's wavelet produces the best results on sharp edges and low-noise smooth areas.

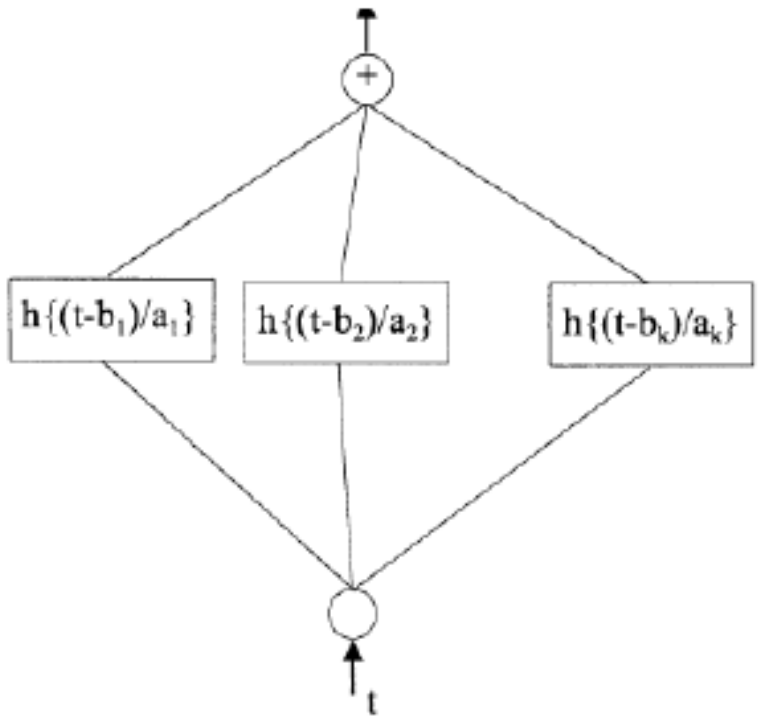

Figure 3 Structure of Wavelet Neural Network

## D. *The Neuro-Wavelet Model*

A neuro-wavelet model [50] for compression of digital images which combines the advantage of wavelet transform and neural network was proposed by Vipula Singh et. al. Here images are decomposed using wavelet filters into a set of sub bands with different resolution corresponding to different frequency bands. Different quantization and coding schemes are used for different sub bands based on their statistical properties. The coefficients in low frequency band are compressed by differential pulse code modulation (DPCM) and the coefficients in higher frequency bands are compressed using neural network. Satisfactory reconstructed images with large compression ratio can be achieved by this method.

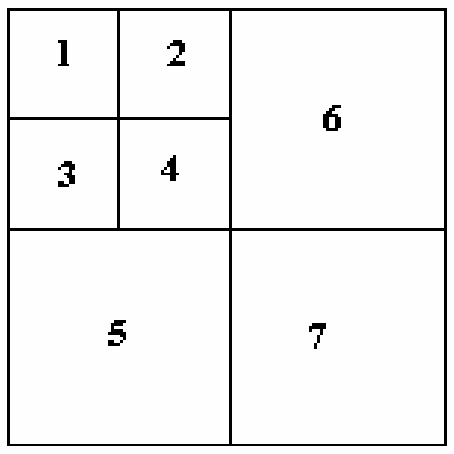

Figure 4 Decomposition on frequency plane by wavelet transform

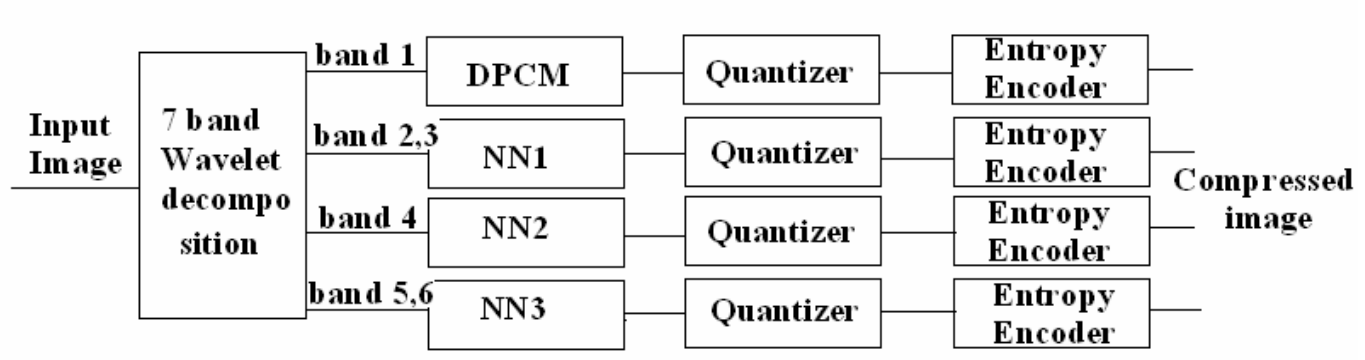

Figure 5 Image compression system based on the neuro-wavelet model

The neuro-wavelet model is described in the fig. 5. First the image is decomposed into different sub bands using wavelet transforms as shown in the fig. 4. Since the human visual system has different sensitivity to different frequency components, the scheme shown here may be adapted. The low frequency band i.e band1 is encoded with DPCM (Differential Pulse Code Modulation). After that these coefficients are scalar quantized. The remaining frequency bands are coded using neural networks. Band2 and Band3 contain the same frequency contents for different orientation. Same neural network is used to compress the data in these bands and a different neural network is used for both band5 and band6. Band4 coefficient is coded using a separate neural network as frequency characteristics of this band does not match with other bands. Band7 is discarded as it contains little information to contribute to the image. This band has little effect on the quality of the reconstructed image. The output of the hidden layer of the neural network is then scalar quantized. Finally these quantized values are entropy coded. Huffmann coding is used here. This paper concludes that compared to the traditional neural network compression or the classical wavelet based compression applied directly on the original image, the neural network based approach improved the quality of the reconstructed image. It also enhances the overall processing time.

TABLE II. BIT RATE & PSNR FOR DIFFERENT WAVELET FILTERS FOR LENA IMAGE

| Wavelet Filter | Bit Rate in bpp (bits per pixel) | PSNR in dB |
|---|---|---|
| ***db18*** | 0.40 | 29.371 |
| ***db6*** | 0.39 | 29.25 |
| ***db4*** | 0.37 | 29.058 |

Experiments were conducted using the images ‘lena’, ‘pepper’, and ‘house’ of size 256 x 256, with $2^8 = 256$ gray levels. Image was decomposed using Daubechies’ 4-coefficient filter (DAUB 4), 6-coefficient filter (DAUB 6) and 18 coefficient filter (DAUB 18). Band-1 is coded using DPCM, Band-2 and 3 is coded using a neural network with eight units in the input and the output and 6 hidden units i.e 8-6-8 neural network. Band-4 is coded using a 8-4-8 neural network, and band-5 and 6 using 16-1-16 network. Scalar quantization and Huffman coding was used on the coefficients of hidden layer. PSNR was evaluated for the resulting image and the comparison between bit rate and PSNR are shown in table 2.

## VI. CONCLUSION

This paper takes a detailed survey on the existing and most significant hybrid methods of image coding. Every approach is found to have its own merits and demerits. VQ based hybrid approaches to compression of the images helps in improving the PSNR along with reducing the computational complexity. It is seen that good quality reconstructed images are obtained, even at low bit-rates when wavelet based hybrid methods are applied to image coding. The powerful parallel processing and learning capability of neural networks can be fully exploited in the ANN based hybrid approaches to image compression. Predictive coding neural networks are found very suitable for compression of text files. High encoding efficiency and good reproduction quality are realized with this type of compression. Fractal neural networks reduce the computation time of encoding/decoding since the compression/ decompression process can be executed in parallel. Simulated results show that the neural network approach can obtain a high compression-ratio and a clear decompressed image. Wavelet networks applied to image compression provide improved efficiency compared to the classical neural networks. By combining wavelet theory and neural networks capability, significant improvements in the performance of the compression algorithm can be realised. Results have shown that the wavelet networks approach succeeded in improving performance and efficiency, especially for rates of compression lower than 75%. The proposed scheme of neuro-wavelets in section C can achieve good compression at low bit rates with good quality reconstructed images. It can be considered that the integration of classical with soft computing based image compression methods enables a new way of achieving higher compression ratio.

The existing conventional image compression technology can be developed by combining high performance coding algorithms in appropriate ways, such that the advantages of both techniques are fully exploited. This will be a significant development and a wide research area in the sense that various traditional image compression algorithms can be empowered with different successful algorithms to achieve better performance. Future research work can be made in this direction to build up new advancements in the field of image compression.

## ACKNOWLEDGMENT

The authors thank the reviewers for their valuable comments and suggestions that helped us to make the paper in its present form.

## AUTHORS PROFILE

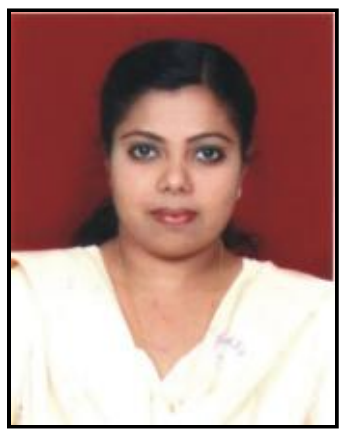

**Rehna. V. J** was born in Trivandrum, Kerala State, India in 1980. She studied Electronics & Communication Engineering at the PET Engineering college, Vallioor, Tirunelveli District, Tamilnadu State, India fom 1999 to 2003. She received Bachelor’s degree from Manonmanium Sundarnar University, Tirunelveli in 2003. She did post graduation in Microwave and TV Engineering at the College of Engineering, Trivandrum and received the Master’s degree from Kerala University, Kerala, India in 2005. Presently, she is a research scholar at the Department of Electronics and Communication Enginering, Noorul Islam Center for Higher Education, Noorul Islam University, Kumarakoil, Tamilnadu, India; working in the area of image processing under the supervision of Dr. M. K. Jeya Kumar.

She is currently working as Assistant Professor at the Department of Telecommunication Engineering, Atria Institute of Technology, Bangalore, India. She has served as faculty in various reputed Engineering colleges in South India over the past seven years. She has presented and published a number of papers in national/international journals/conferences. She is a member of the International Association of Computer Science & Information Technology (IACSIT) since 2009. Her research interests include numerical computation, soft computing, enhancement, coding and their applications in image processing.

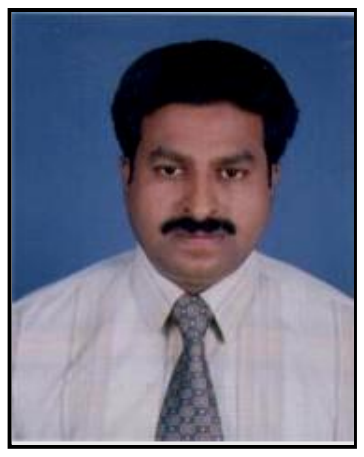

**M. K. Jeya Kumar** received his PhD degree in Mobile Adhoc Networks from Dr. MGR University, Chennai, India, in 2010. He is Assistant Professor at the Department of Computer Application, Noorul Islam University, Kanyakumari District, Tamilnadu, India. His research interests include networks and network security, image processing and soft computing techniques.